\documentclass{iopart}

\usepackage[dvips]{graphicx}

\begin{document}

\title{Charmonium-hadron interactions from QCD}

\author{Su Houng Lee and Yongseok Oh}

\address{Institute of Physics and Applied Physics,
Department of Physics, Yonsei University, Seoul 120-749, Korea}

\ead{\mailto{suhoung@phya.yonsei.ac.kr}, \mailto{yoh@phya.yonsei.ac.kr}}

\begin{abstract}
The heavy quark system is an excellent probe to learn about the QCD
dynamics at finite density.
First, we discuss the properties of the $J/\psi$ and $D$ meson
at finite nucleon density.
We discuss why their properties should change at finite density and
then introduce an exact QCD relation among these hadron properties and
the energy momentum tensor of the medium.
Second, we discuss attempts to calculate charmonium-hadron total cross
section using effective hadronic models and perturbative QCD.
We emphasize a recent calculation, where the cross section is
derived using QCD factorization theorem.
We conclude by discussing some challenges for SIS 200.
\end{abstract}

%\newpage

\section{ $J/\psi$ and $D$ mesons at finite density}

In the heavy quark system, the heavy quark mass $m_H^{}$ provides a
natural normalization scale $\mu$, which makes the perturbative QCD
approach possible even for calculating some bound state properties
of the heavy quarks.
This implies that the heavy quark system is an excellent probe to learn
about the QCD dynamics at finite density and/or temperature.
Here, we will highlight some calculations which explicitly demonstrate
the relations between the changes of the properties of the heavy quark
system and the QCD dynamics at finite density.

\subsection{ $J/\psi$ mass at finite density}

Let us first start with the $J/\psi$ meson mass at finite density.
The interaction between the $J/\psi$ and a hadron $h$ can be
perturbatively generated by multiple gluon exchanges.
This implies that the mass shift of the $J/\psi$ in nuclear medium
can be estimated if the gluon distribution is known in nuclear
medium \cite{Bro90}.
In fact, in the heavy quark mass limit $m_c \rightarrow \infty$, the
mass shift can be calculated exactly to leading order in QCD \cite{Luk92}
and is given as
\begin{eqnarray}
\Delta m_{J/\psi}  = 
 c_1 \times \Delta \langle E^2 \rangle_{\rm matter},
\label{jpsimass}
\end{eqnarray}
where $c_1$ is a calculable constant and the matrix element is taken
with respect to the nuclear matter.
To the leading order in density, the matrix element is known from the
trace anomaly relation and the gluon distribution function of the nucleon
\cite{VZ80}.  
For a more realistic charm quark mass, there are some model dependence on 
how to treat the bound state part.
So far, there are potential model approaches \cite{Bro90,Was91} and QCD
sum rule approaches \cite{Kli99,KL01}.
In all cases, similar relation to Eq.~(\ref{jpsimass}) holds with
slightly different expressions for $c_1$, such that the mass shift at
nuclear matter density ranges from $-4$ to $-7$ MeV.
The mass shift is small at nuclear matter density and one doubts whether
such a small mass shift can be observed.
However, one can turn the argument around and claim that if any mass shift
for $J/\psi$ is observed, it will tell us about the changes of gluon
field configuration at finite density through Eq.~(\ref{jpsimass}).

\subsection{ $D$ meson mass at finite density}

In the $D$ meson, the heavy quark acts as a source for the light quark,
which surrounds the heavy quark and probes the QCD vacuum.
This is the basic picture of the constituent quark in the  heavy-light
meson system within the formulation based on the heavy quark
symmetry \cite{Wisgur}. 
Therefore, if the vacuum properties are changed at finite temperature or 
density, the light quark should be sensitively affected and be reflected 
in the changes of the $D$ meson properties in matter.
For the $D$ meson, there does not exist a formal limit where one can 
calculate the mass change in matter.
Nevertheless, model calculations suggest \cite{AE83,TTS99,H00} that it
is dominantly related to the change in the chiral condensate.
In fact, in both the quark-meson coupling model \cite{TTS99} and in the
QCD sum rule approach \cite{H00}, the average mass shift of the $D^\pm$
mesons at normal nuclear matter density were found to be around $-50$ MeV.

The $D$ meson mass shift has several importance in the $J/\psi$
suppression phenomena in relativistic heavy ion collisions (RHIC).
First, it leads to subthreshold production and the change of the $D\bar{D}$
threshold for the $J/\psi$ system in matter \cite{STST00}.
Second, a change of the averaged $D^{\pm}$ mass might lead to level
crossings of the $\bar{D}D$ threshold with the $\psi'$  and $\chi$ 
mass \cite{H00,DPS01}, which leads to a step-wise $J/\psi$ suppression
\cite{H00,DPS011}.
Unfortunately, for the $D$ meson mass shift, there are large model
dependence and nontrivial splitting between the $D^+$ and $D^-$ mesons
\cite{MLW}.

\subsection{Exact sum rule from QCD}

The $D$ meson mass and the $D\bar{D}$ threshold have important
phenomenological consequences.
Unfortunately, only model-dependent calculation could be made.
Here, we will introduce an exact QCD equality, which can be used to link
the $D\bar{D}$ threshold to $J/\psi$ suppression.
The exact sum rules in QCD at finite temperature and/or density were 
first introduced to the correlation functions between light hadrons
\cite{HL95} and the derivation goes as follows.
Let us consider the correlation function between two meson currents
and its dispersion relation,
\begin{eqnarray}
\Delta \Pi(Q^2,T,\rho) & = & i \int d^4x e^{iqx} \Delta
\langle \bar{q} \Gamma q(x) \bar{q}\Gamma q(0) \rangle
\nonumber \\
 & = & \frac{1}{\pi}\int ds \frac{\Delta {\rm Im} \Pi(s)}{s+Q^2}.
\label{correlation}
\end{eqnarray}
Here, we have assumed $q_\mu=(iQ,0,0,0)$ and $\Delta \langle \cdot
\rangle= \langle \cdot \rangle_{\rm medium} -  \langle \cdot
\rangle_{\rm vacuum}$ means the difference between the medium and the
vacuum expectation values.
$\Gamma$ is a Dirac gamma matrix that will be determined by what
meson we want to study.
At $Q^2 \rightarrow \infty$, the real part of Eq.~(\ref{correlation})
can be determined from the operator product expansion (OPE) in QCD,
\begin{eqnarray}
\Delta \Pi(Q^2,T,\rho)=\sum_n \frac{1}{(Q^2)^{d_n-2}} \bigg(
\frac{g^2(Q)}{g^2(\mu)} \bigg)^{\frac{2 \gamma_J-\gamma_n}{2b}}
C_n \Delta \langle O_n(\mu) \rangle.
\label{anomalous}
\end{eqnarray}
Here, $\gamma_n$ $ (\gamma_J)$ is the anomalous dimension of the
operator (current).
Equation (\ref{anomalous}) is an asymptotic expansion in $1/\log(Q^2)$
and $1/Q^2$.
For $\gamma_n=\gamma_J=0$, the leading term is proportional to $1/Q^2$
and can be related to the first moment of the spectral density as
\cite{HL95}
\begin{eqnarray}
\int ds \frac{s}{\pi} \Delta {\rm Im}\Pi(s)= 
c_4  \Delta \langle O_4 \rangle,
\label{leading}
\end{eqnarray}
where $O_4$ is a dimension 4 operator and $c_4$ its corresponding 
Wilson coefficient.

\begin{figure}
\begin{center}
\includegraphics[height=4cm]{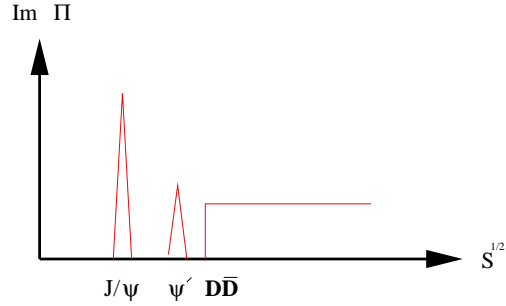}
\caption{\label{fig:sfig3} Charmonium spectral density.}
\end{center}
\end{figure}

One can generalize this formula for the heavy (charm) vector current 
$\bar{c} \gamma_\mu c $ at rest and obtain the following constraint:
\begin{eqnarray}
\int ds \frac{s}{\pi} \Delta {\rm Im}\Pi(s)= 
-\frac{1}{12}  \Delta \langle \frac{\alpha_s}{\pi}G^2 \rangle 
+\frac{8}{16+3n_f} \Delta \langle T_{00} \rangle,    
\end{eqnarray}
where $T_{\mu \nu}$ is the traceless energy momentum tensor and
$\langle \frac{\alpha_s}{\pi}G^2 \rangle$ is the gluon condensate,
which is related to the trace of the energy momentum tensor.
Lattice gauge theory provides the temperature dependence of the 
operators both below and above the phase transition temperature.
From this, one can investigate what kind of changes in the spectral
density, which in the vacuum has a generic form given in
Fig.~\ref{fig:sfig3}, is consistent with the constraint equation. 

Now let us look at the changes at finite density.
Using the linear density approximation $\Delta \langle \cdot \rangle =
\frac{\rho}{2m} \langle p | \cdot | p \rangle$, one has
\begin{eqnarray}
-\frac{1}{12}  \Delta \langle \frac{\alpha_s}{\pi}G^2 \rangle 
&\approx& 6.25 \times 10^{-5} \times x \mbox{ GeV}^4,  \\ \nonumber  
\frac{8}{16+3n_f} \Delta \langle T_{00} \rangle &\approx& 3 
\times 10^{-4} \times x \mbox{ GeV}^4, 
\label{order}     
\end{eqnarray}
where $x=\rho/\rho_{\rm n.m.}$ with $\rho_{\rm n.m.}=0.17/(\mbox{fm})^3$. 
In comparison with this, when neglecting the contribution from the $\psi'$, 
the spectral density in Fig.~\ref{fig:sfig3} can be parameterized as
\begin{eqnarray}
\frac{1}{\pi} {\rm Im}\Pi(s)= f_V^2 \delta (s-m_{J/\psi}^2) + 
\frac{1}{4\pi^2}(1+ \frac{\alpha_s}{4\pi}) \theta(s-s_0),
\end{eqnarray}
where $f_V^2=\frac{3 m_{J/\psi}}{4 \pi \alpha_e^2 Q_c^2} 
\Gamma(J/\psi \rightarrow e^+e^-)\approx 0.16 $ GeV$^2$.
Hence, we have
\begin{eqnarray}
\int ds \frac{s}{\pi} \Delta {\rm Im}\Pi(s)= 
\Delta (f_V^2 m_{J/\psi}^2) -
\frac{1}{8\pi^2}(1+ \frac{\alpha_s}{4\pi}) \Delta (s_0^2).
\label{phen}
\end{eqnarray}
Now, since the mass shift $\Delta m_{J/\psi}$ is very small, the change
in the first term is coming from the change in $f_V^2$, which is related
to the change of the coupling of the $J/\psi$ to the electromagnetic current.
This is precisely related to the amount of $J/\psi$ suppression in matter.
At linear density approximation, as can be seen from Eq. (6),
the expected change in the OPE is of $10^{-4}$ GeV$^4$ in the nuclear
matter and can be neglected compared to a fractional change in the
phenomenological side of Eq. (\ref{phen}).
Therefore, in this case, the suppression in $f_V^2$ is related to the
decrease of $s_0$.
In an order of magnitude estimate, we arrive at
\begin{eqnarray}
\frac{1}{2} \Delta s_0^{1/2} ={ m_{J/\psi}^2 f_V^2 \pi^2 \over
(1+\frac{\alpha_s}{4\pi}) s_0^{3/2}} \times x 
 \sim 280 \times  { \Delta f_V^2 \over f_V^2 } \mbox{ MeV},
\label{approx}
\end{eqnarray}
where ${\Delta f_V^2}/{f_V^2}$  quantifies the amount of $J/\psi$ 
suppression seen through the dilepton signal.
Assuming that $s_0^{1/2}= m_D+ m_{\bar{D}}$, Eq. (\ref{approx}) implies
that 10\% suppression is related to 28 MeV mass decrease of the $D$
meson in nuclear matter.

The relation in Eq. (\ref{leading}) can be used in the other ways.
Namely, from the measurement of the changes in the mass of the $D$
mesons and the suppression of the $J/\psi$ peak, one is able to learn
about the changes in the energy momentum tensor and gluon condensate
at finite temperature or density, where the measurement would be made.

\section{$J/\psi$-hadron cross section}

\begin{figure}
\begin{center}
\includegraphics[height=8cm,angle=-90]{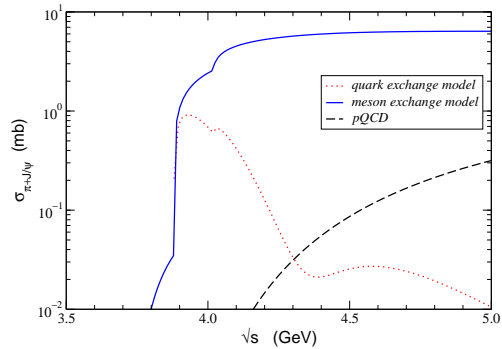}
\caption{\label{fig:com} $\sigma_{J/\psi+\pi}$ cross section from meson
exchange model \cite{meson-ex}, quark exchange model \cite{WSB01} and
perturbative QCD \cite{BP79} calculations.}
\end{center}
\end{figure}

$J/\psi$ suppression \cite{Matsui:1986dk} seems to be one of the
most promising signal for QGP formation in RHIC.
Indeed the recent data at CERN \cite{Abreu:2000ni} show an anomalous
suppression of $J/\psi$ formation, which seems to be a consequence of QGP 
formation \cite{Blaizot:2000ev}.
However, before coming to a concrete conclusion, one has to estimate the
amount of $J/\psi$ suppression due to other non-QGP mechanism, one of
which is the hadronic final state interactions.
Indeed, there are model calculations that show large suppression by
hadronic final state interactions \cite{capella99,capella00,Ko97}.
The typical values for $J/\psi + \mbox{hadron}$ dissociation
cross sections one uses for these calculations are
$\sigma_{J/\psi+N} \sim 5 $ mb and $\sigma_{J/\psi+comover} \sim 1 $ mb.
However, there is no direct experimental data from which we can determine
or confirm these cross sections.
In fact, the existing model calculations vary greatly in their energy
dependence and magnitude near the threshold.
In Fig.~\ref{fig:com}, we show the latest typical results of model
calculations based on perturbative QCD
\cite{Peskin79,BP79,KS94,AGGA01,OKL01}, the meson exchange model
\cite{meson-ex} and the quark exchange model \cite{WSB01}.    
To understand the discrepancies and to obtain a consistent result,
it is crucial to probe each model calculations further so as to
spell out the limitations and range of validity of each model
calculations.
Here, we will have a closer look at the perturbative QCD result.

\subsection{Perturbative QCD methods}

More than two decades ago, Peskin \cite{Peskin79} and Bhanot and Peskin 
\cite{BP79} showed that one could apply perturbative Quantum
Chromodynamics (pQCD) to calculate the interactions between the bound
state of heavy quarks and the light hadrons.
Such calculation was feasible because in the large quark mass limit one
could consistently obtain the leading order OPE of the correlation
function between two heavy meson currents in the light hadron state.
The justification of such calculation stems from the fact that there exist
two relevant scales in the bound state \cite{ADM78} in the large quark
mass limit, namely the binding energy, which becomes Coulomb-like and
scales as $mg^4$, and the typical momentum scale of the bound state,
which scales like $mg^2$, where $m$ is the heavy quark mass and $g$ the
quark-gluon coupling constant.
Hence, taking the separation scale of the OPE to be the binding
energy, it is consistent to take into account the bound state
property, which is generated by the typical momentum scale of the
bound state, into the process-dependent Wilson coefficient.
The result obtained by Peskin and Bhanot was derived within
the OPE \cite{Peskin79,BP79}.
Recently, we have derived anew the leading order pQCD result using
the QCD factorization theorem \cite{OKL01}.
The equivalence between the OPE and the factorization approach is well
established in the deep inelastic scattering and Drell-Yan processes
\cite{HVN81}.
Our work shows that the equivalence between the two approaches are
also true for bound state scattering to leading order in QCD.
Similar approach has been used by one of us (SHL) to estimate the
dissociation cross section of the $J/\psi$ at finite temperature
\cite{HLZ88}.
The factorization formula also provides a manageable starting point to
calculate higher twist gluonic effects \cite{BB00,KL01}, which should be
nontrivial for the $J/\psi$-hadron scattering.
Here, we will sketch the derivation .

We refer to a bound state of heavy quark and its own antiquark as $\Phi$.
According to the factorization formula, the total scattering cross section
of $\Phi$ with a hadron $h$ can be written as \cite{BP79}
\begin{eqnarray}
\sigma_{\Phi h}(\nu) = \int_0^1 dx \sigma_{\Phi g}(x \nu) g(x),
\label{eq:fact}
\end{eqnarray}
with $\nu = p \cdot q / M_\Phi$, where $p$ ($q$) is the momentum of the
hadron ($\Phi$) and $M_\Phi$ is the $\Phi$ mass.
Here, $\sigma_{\Phi g}$ is the perturbative $\Phi$-gluon scattering
cross section and $g(x)$ is the leading twist gluon distribution function
within the hadron.
The separation scale is taken to be the binding energy of the bound state.
Hence, the bound state properties have to be taken into account in
$\sigma_{\Phi g}$.
This can be accomplished by introducing the Bethe-Salpeter (BS) amplitude
$\Gamma(p_1,-p_2)$, which  satisfies \cite{BS51}
\begin{eqnarray}
\Gamma_\mu(p_1,-p_2) &=& i C_{\rm color} \int \frac{d^4 k}{(2\pi)^4} g^2
V(k) \gamma^\nu \Delta(p_1+k)
\nonumber \\ && \mbox{} \times
\Gamma_\mu(p_1+k,-p_2+k) \Delta(-p_2+k)
\gamma_\nu,
\nonumber \\
\end{eqnarray}
where $C_{\rm color} = (N_c^2-1)/(2N_c)$ with the number of color $N_c$.
The trivial color indices have been suppressed and $p_1$ ($-p_2$) is the
four-momentum of the heavy quark (anti-quark).

We write $q=p_1+p_2$ and $p = (p_1-p_2)/2$ and work in the $\Phi$
rest frame and introduce the binding energy $\varepsilon$, such that
$q_0 = M_\Phi = 2 m+ \varepsilon$  ($\varepsilon<0$).  
Then, in the non-relativistic limit, the BS amplitude reduces to
the following form:
\begin{eqnarray}
\Gamma_\mu({\textstyle\frac12}q+p,-{\textstyle\frac12}q+p)
&=& - \left( \varepsilon - \frac{{\bf p}^2}{m} \right)
\sqrt{\frac{M_\Phi}{N_c}}
%\nonumber \\ &&
\psi({\bf p})
\frac{1+\gamma_0}{2} \gamma_i \delta_{\mu i} \frac{1-\gamma_0}{2}, 
\nonumber \\ &&
\label{eq:BS}
\end{eqnarray}
and the corresponding BS equation becomes the non-relativistic
Schr\"odinger equation for the Coulombic bound state,
so that $\psi({\bf p})$ is the normalized wave function for the bound state.  

\begin{figure}
\begin{center}
\includegraphics[height=3.5cm]{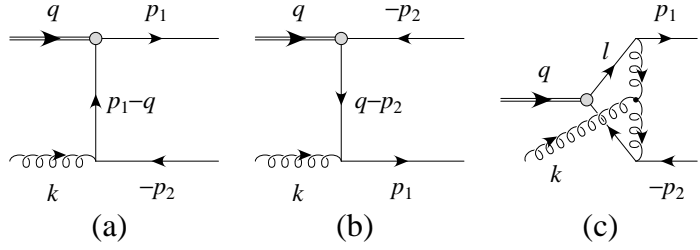}
\caption{\label{figreac} Scattering processes of $\Phi$ and gluon.}
\end{center}
\end{figure}

With the BS amplitude defined as Eq. (\ref{eq:BS}), we now obtain the
$\Phi$-gluon scattering amplitude with the processes depicted in
Fig.~\ref{figreac}.
The scattering matrix elements can be obtained by substituting the 
BS amplitude for the vertex.
To obtain the leading order result, we should pay attention to the
counting scheme.
First the binding energy $\varepsilon_0 = m \left[ N_c g^2 / (16\pi)
\right]^2$ is $O(mg^4)$.
Combined with the energy conservation $q+k=p_1+p_2$ in the non-relativistic
limit, this implies $|{\bf p_1}| \sim |{\bf p_2}| \sim O(mg^2)$ and
$k^0=|{\bf k}|\sim O(mg^4)$.
Within this counting scheme, the final answer reads
\begin{eqnarray}
\mathcal{M}^{\mu\nu} &=& -g \sqrt{\frac{M_\Phi}{N_c}}
\left\{ {\bf k} \cdot \frac{\partial
\psi({\bf p})}{\partial {\bf p}} \delta^{\nu 0} + k_0 \frac{\partial
\psi({\bf p})}{\partial p^j} \delta^{\nu j} \right\} \delta^{\mu i}
\nonumber \\ && \mbox{} \times
\bar{u}(p_1) \frac{1+\gamma_0}{2} \gamma^i \frac{1-\gamma_0}{2}T^a v(p_2).
\label{eq:amp}
\end{eqnarray}

The scattering cross section $\sigma_{\Phi g}$ can now be obtained from 

\begin{eqnarray}
\sigma_{\Phi g} =\int \frac{1}{4 M_\Phi k_0}
\overline{|\mathcal{M}|^2}
(2\pi)^4 \delta^4(p_1+p_2-k-q) \frac{d^3 p_1d^3 p_2}{4p_1^0 p_2^0 (2\pi)^6}.
\label{phase}
\end{eqnarray}
With the amplitude (\ref{eq:amp}) we obtain
\begin{equation}
\overline{|\mathcal{M}|^2} = \frac{4g^2 m^2 M_\Phi k_0^2}{3 N_c} \left|
{\bf \nabla} \psi ({\bf p}) \right|^2.
\label{eq:Mfin2}
\end{equation}
Substituting Eq. (\ref{phase}) into Eq. (\ref{eq:fact}) and using the 
known gluon distribution function in the pion, we obtain the dashed line
in Fig.~\ref{fig:com}.

\begin{figure}
\begin{center}
\includegraphics[height=4.5cm]{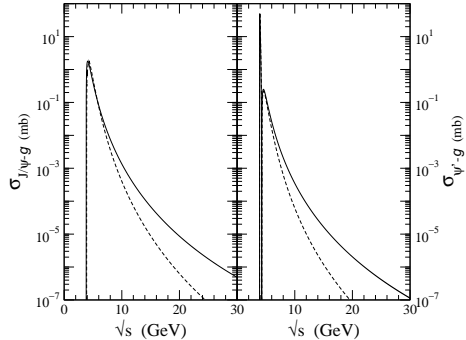}
\caption{\label{fig:elem} Scattering cross sections $\sigma_{J/\psi-g}$
and $\sigma_{\psi'(2S)-g}$.
The solid lines are obtained in the non-relativistic limit and the
dashed lines are with the relativistic correction.}
\end{center}
\end{figure}

\subsection{Corrections to leading order formula}

The counting of this formalism implies that for the charmonium system
$\alpha_s=0.84$ and the separation scale $\varepsilon_0=0.78$ GeV.
Hence, both the $\alpha_s$ and higher twist effects should be important.
However, calculating higher $\alpha_s$ corrections in this formalism is
intricately related to the relativistic corrections including $1/m$
corrections, which is a formidable future task.

Here, we will only look at the simple but important relativistic correction
coming from the relativistic calculation of the phase space integral in
Eq.~(\ref{phase}).
The difference after a full calculation is shown in Fig.~\ref{fig:elem}
for $\sigma_{\Phi g}$, which shows larger suppression of the cross
sections at higher energy.
However, we find that this relativistic correction has almost no effect
on the cross section $\sigma_{\Phi h}$.  This follows because 
$\sigma_{\Phi g}$ is highly peaked at small energy region and the 
gluon distribution function increases towards small $x$.  
Through Eq.~(\ref{eq:fact}), this implies that 
$\sigma_{\Phi h}$, at all energies, are dominated by the low energy 
behavior of $\sigma_{\Phi g}$, which is reliably calculated in the 
non-relativistic limit.  
This implies that the full relativistic corrections to the scattering
amplitude $\sigma_{\Phi g}$ would have little effects on $\sigma_{\Phi h}$.

The remaining important and interesting correction is the higher
twist-effect, which contributes as $A/\varepsilon^2$ in our formalism.
Since $A$ is related to some hadronic scale of the hadron $h$, it should
be of the order $\varepsilon^2$ itself, hence should be 
non-negligible \cite{Ge98}.
It is also important to extend this calculation to investigate the absortion 
cross section for the $\chi$ states.
This will influence the amount of $J/\psi$ production 
coming from the decay of the $\chi$'s
in a p-A or A-A reaction \cite{Ge98}.

\section{Conclusion and challenges for SIS200}

As we have discussed, observation of the changes in the properties of
the heavy quark system will provide us useful information about the QCD
dynamics both in the vacuum and in the medium. 
Important experimental observation would be mass shifts of the charmonium
and open-charm system in nuclear medium.
Also important would be the quantitative observation of the suppression
of the $J/\psi$ and $\psi'$ signals in the medium through the dileptons.

$\bar{p}$-$A$ experiments to probe the $J/\psi$ in nuclear matter
would also provide useful informations on the $J/\psi$-nucleon cross
section at low energy.
This is precisely where there is huge discrepancy between existing
theoretical predictions. 
Experimental investigation will provide useful information on the
higher twist effects in the pQCD result for the
$J/\psi$-nucleon cross section.
Here, the higher twist effect comes solely from gluon operators and
hence this in turn would be a unique place to learn about the higher
twist gluon effect of the nucleon and nuclear matter.

\ack

This work was supported in part by the Brain Korea 21 project of Korean
Ministry of Education, KOSEF Grant 1999-2-111-005-5, the
Yonsei University Research Grant, and the Korean Ministry of Education
under Grant 2000-2-0689.

\section*{References}
%\begin{harvard}

%\end{harvard}
%\endrefs


\begin{thebibliography}{99}

\bibitem{Bro90}
Brodsky S J, Schmidt I A and de Teramond G F 1990
{\it Phys. Rev. Lett.} {\bf 64} 1011
%%CITATION = PRLTA,64,1011;%%

\bibitem{Luk92}
Luke M E, Manohar A V and Savage M J 1992
{\it Phys. Lett.} B {\bf 288} 355
%%CITATION = PHLTA,B288,355;%%

\bibitem{VZ80}
Voloshin M and Zakharov V 1980
{\it Phys. Rev. Lett.} {\bf 45} 688
%%CITATION = PRLTA,45,688;%%

\bibitem{Was91}
Wasson D A 1991
{\it Phys. Rev. Lett.} {\bf 67} 2237
%%CITATION = PRLTA,67,2237;%%

\bibitem{Kli99}
Klingl F, Kim S, Lee S H, Morath P and Weise W 1999
{\it Phys. Rev. Lett.} {\bf 82} 3396
%%CITATION = PRLTA,82,3396;%%

\bibitem{KL01}
Kim S and Lee S H 2001
{\it Nucl. Phys.} A {\bf 679} 517
%%CITATION = NUPHA,A679,517;%%

\bibitem{Wisgur}
Isgur N and Wise M B 1989
{\it Phys. Lett.} B {\bf 232} 113 \\
%%CITATION = PHLTA,B232,113;%%
Isgur N and Wise M B 1990
{\it Phys. Lett.} B {\bf 237} 527 
%%CITATION = PHLTA,B237,527;%%
 
\bibitem{AE83}
Aliev T M and Eletsky V L 1983
{\it Sov. J. Nucl. Phys.} {\bf 38} 936  
%%CITATION = SJNCA,38,936;%%

\bibitem{TTS99}
Tsushima K \etal 1999
{\it Phys. Rev. } C {\bf 59} 2824
%%CITATION = PHRVA,C59,2824;%%

\bibitem{H00}
Hayashigaki A 2000
{\it Phys. Lett.} B {\bf 487} 96
%%CITATION = PHLTA,B487,96;%%

\bibitem{STST00}
Sibirtsev A \etal 2000
{\it Phys. Lett. } B {\bf 484} 23  
%%CITATION = PHLTA,B484,23;%%

\bibitem{DPS01}
Digal S, Petreczky P and Satz H 2001
{\it Phys. Lett. } B {\bf 514} 57
%%CITATION = PHLTA,B514,57;%%

\bibitem{DPS011}
Digal S, Petreczky P and Satz H 2001
{\it Phys. Rev. } D {\bf 64} 094015
%%CITATION = PHRVA,C64,094015;%%

\bibitem{MLW} 
Morath P, Weise W and Lee S H 1999
{\it Autumn School in Physics, QCD: Perturbative or Nonperturbative?}
(Singapore: World Scientific) p 425

\bibitem{HL95}
Huang S-Z and Lissia M 1995
{\it  Phys. Lett.} B {\bf 348} 571 \\
%%CITATION = PHLTA,B348,571;%%
Huang S-Z and Lissia M 1995
{\it  Phys. Rev. } D {\bf 52} 1134  
%%CITATION = PHRVA,D52,1134;%%

%\bibitem{LF01}
%Lee S H and Friman B
%in preparation

\bibitem{Matsui:1986dk}
Matsui T and Satz H 1986
{\it Phys. Lett.}  B {\bf 178} 416
%%CITATION = PHLTA,B178,416;%%

\bibitem{Abreu:2000ni}
Abreu M C \etal  [NA50 Collaboration] 2000
{\it Phys. Lett.} B {\bf 477} 28
%%CITATION = PHLTA,B477,28;%%

\bibitem{Blaizot:2000ev}
Blaizot J-P, Dinh M and Ollitrault J-Y 2000
{\it Phys. Rev. Lett.} {\bf 85} 4012
%%CITATION = PRLTA,85,4012;%%

\bibitem{capella99}
Armesto N, Capella A and Ferreiro E G 1999
{\it Phys. Rev.} C {\bf 59} 395
%%CITATION = PHRVA,C59,395;%%

\bibitem{capella00}
Capella A, Ferreiro E G and Kaidalov A B 2000
{\it Phys. Rev. Lett.} {\bf 85} 2080
%%CITATION = PRLTA,85,2080;%%

\bibitem{Ko97}
Cassing W and Ko C M 1997
{\it Phys. Lett.} B {\bf 396} 39
%%CITATION = PHLTA,B396,39;%%

\bibitem{Peskin79}
Peskin M E 1979
{\it Nucl. Phys.} {\bf B156} 365
%%CITATION = NUPHA,B156,365;%%

\bibitem{BP79}
Bhanot G and Peskin M E 1979
{\it Nucl. Phys.} {\bf B156} 391
%%CITATION = NUPHA,B156,391;%%

\bibitem{KS94}
Kharzeev D and Satz H 1994
{\it Phys. Lett.} B {\bf 334} 155\\
%%CITATION = PHLTA,B334,155;%%
Kharzeev D, Satz H, Syamtomov A and Zinovjev G 1996
{\it Phys. Lett.} B {\bf 389} 595
%%CITATION = PHLTA,B389,595;%%

\bibitem{AGGA01}
Arleo F, Gossiaux P-B, Gousset T and Aichelin J 2002
{\it Phys. Rev.} D {\bf 65} 014005
%%CITATION = PHRVA,D65,014005;%%
   
\bibitem{OKL01}
Oh Y, Kim S and Lee S H 2001
{\it Preprint} hep-ph/0111132
%%CITATION = HEP-PH 0111132;%%

\bibitem{meson-ex}
Matinyan S G and M{\"u}ller B 1998
   {\it Phys. Rev.} C {\bf 58} 2994 \\
%%CITATION = PHRVA,C58,2994;%%
Haglin K L 2000
   {\it Phys. Rev.} {\bf 61} 031902 \\
%%CITATION = PHRVA,C61,031902;%%
Lin Z and Ko C M 2000
   {\it Phys. Rev.} {\bf 62} 034903 \\
%%CITATION = PHRVA,C62,034903;%%
Oh Y, Song T and Lee S H 2001
   {\it Phys. Rev.} {\bf 63} 034901 \\
%%CITATION = PHRVA,C63,034901;%%
Sibirtsev A, Tsushima K and Thomas A W 2001
   {\it Phys. Rev.} {\bf 63} 044906 \\
%%CITATION = PHRVA,C63,044906;%%
Liu W, Ko C M and Lin Z W 2001
   {\it Preprint} nucl-th/0107058
%%CITATION = NUCL-TH 0107058;%%

\bibitem{WSB01}
Wong C-Y, Swanson E S and Barnes T E 2000
   {\it Phys. Rev.} C {\bf 62} 045201 \\
%%CITATION = PHRVA,C62,045201;%%
Wong C-Y, Swanson E S and Barnes T E 2002
   {\it Phys. Rev.} C {\bf 65} 014903 \\
%%CITATION = PHRVA,C65,014903;%%
Wong C-Y 2001
   {\it Preprint} nucl-th/0110004
%%CITATION = NUCL-TH 0110004;%%

\bibitem{ADM78}
Appelquist T, Dine M and Muzinich I 1978
   {\it Phys. Rev.} D {\bf 17} 2074
%%CITATION = PHRVA,D17,2074;%%

\bibitem{HVN81}
Humpert B and van Neerven W L 1981
   {\it Phys. Lett.} B {\bf 102} 426 \\
%%CITATION = PHLTA,B102,426;%%
Humpert B and van Neerven W L 1982
   {\it Phys. Rev.} D {\bf 25} 2593
%%CITATION = PHRVA,D25,2593;%%

\bibitem{HLZ88}
Hansson T H, Lee Su H and Zahed I 1988
  {\it Phys. Rev.} {\bf D37} 2672
%%CITATION = PHRVA,D37,2672;%%

\bibitem{BB00}
Bartels J, Bontus C and Spiesberger H 1999
   {\it Preprint} hep-ph/9908411 \\
%%CITATION = HEP-PH 9908411;%%
Bartels J and Bontus C 2000
   {\it Phys. Rev.} D {\bf 61} 034009
%%CITATION = PHRVA,D61,034009;%%

\bibitem{BS51}
Salpeter E E and Bethe H A 1951
   {\it Phys. Rev.} {\bf 84} 1232
%%CITATION = PHRVA,84,1232;%%

\bibitem{Ge98}
Gerland L, Frankfurt L, Strikman M, St\"oker H and Greiner W 1998
   {\it Phys. Rev. Lett.} {\bf 81} 762
%%CITATION = PRLTA,81,762;%%

\end{thebibliography}
\end{document}